\documentclass{elsart}

\usepackage{lineno}
\usepackage{graphicx}
\usepackage{amssymb}
\usepackage{color}
\usepackage[justification=centering]{caption}

\def \esym{$E_{\rm sym}(\rho)$}
\def \amev{MeV/u}
\def \arau{$^{40}$Ar+$^{197}$Au}
\def \esym{$E_{\rm sym}(\rho)$}

\begin{document}

\begin{frontmatter}

\title{The Emission Order of Hydrogen Isotopes via Correlation Functions in 30 MeV/u Ar+Au Reactions}

\author[THU]{Yijie Wang},
\author[THU]{Fenhai Guan},
\author[THU]{Qianghua Wu},
\author[THU]{Xinyue Diao},
\author[THU]{Yan Huang},
\author[THU]{Liming Lyu},
\author[THU]{Yuhao Qin},
\author[THU]{Zhi Qin},
\author[BHU]{Dawei Si},
\author[IMP]{Zhen Bai},
\author[IMP]{Fangfang Duan},
\author[IMP]{Limin Duan},
\author[IMP,UCAS]{Zhihao Gao},
\author[IMP]{Qiang Hu},
\author[IMP]{Rongjiang Hu},
\author[IMP]{Genming Jin},
\author[IMP,UCAS]{Shuya Jin},
\author[IMP,UCAS]{Junbing Ma},
\author[IMP]{Peng Ma},
\author[HZU,IMP]{Jiansong Wang},
\author[IMP,UCAS]{Peng Wang},
\author[IMP,UCAS]{Yufeng Wang},
\author[IMP,UCAS]{Xianglun Wei},
\author[IMP]{Herun Yang},
\author[IMP]{Yanyun Yang},
\author[IMP,UCAS]{Gongming Yu},
\author[IMP,UCAS]{Yuechao Yu},
\author[IMP]{Yapeng Zhang},
\author[IMP,UCAS]{Qingwu Zhou},
\author[BNU]{Yaofeng Zhang},
\author[HNU]{Chunwang Ma},
\author[UCAS,SINAP]{Xinrong Hu},
\author[SINAP,SARI]{Hongwei Wang},
\author[ANU]{Yunyi Cui},
\author[ANU]{Junlong Tian},
\author[THU]{Zhigang Xiao\thanksref{info}}

\address[THU]{Department of Physics, Tsinghua University, Beijing 100084, China;}
\address[BHU]{School of Physics, Beihang University, Beijing 100191, China;}
\address[UCAS]{University of Chinese Academy of Sciences, Beijing 100049, China;}
\address[IMP]{Institute of Modern Physics, Chinese Academy of Sciences, Lanzhou 730000, China;}
\address[HZU]{School of Science, Huzhou University, Huzhou, 313000, China;}
\address[BNU]{College of Nuclear Science and Technology, Beijing Normal University, Beijing 100875, China;}
\address[HNU]{Institute of Particle and Nuclear Physics, Henan Normal University, Xinxiang 453007, China;}
\address[SINAP]{Shanghai Institute of Applied Physics, Chinese Academy of Science, Shanghai 201800, China;}
\address[SARI]{Shanghai Advanced Research Institute, Chinese Academy of Science, Shanghai 201210, China;}
\address[ANU]{School of Physics and Electrical Engineering, Anyang Normal University, Anyang 455000, China}

\thanks[info]{E-mail:~xiaozg@tsinghua.edu.cn (corr. author)}

\end{frontmatter}
\begin{frontmatter}

\begin{abstract}

The intensity interferometry is applied as a chronometer of the particle emission of hydrogen isotopes from the intermediate velocity source formed in  \arau~ reactions at 30 \amev. The dynamic emission order of  $\tau_{\rm p}>\tau_{\rm d}>\tau_{\rm t}$ is evidenced  via the correlation functions of nonidentical particle pairs. Assuming the similar source size, the same emission order is inferred from  the correlation functions of  identical particle pairs, where $\tau_{\rm p} \approx 100 {\rm ~fm/c}$ is extracted by the fit of Koonin-Pratt equation to p-p correlation function. Transport model simulations demonstrate that the dynamic emission order of light charged particles depends on the stiffness of the nuclear symmetry energy. 

\end{abstract}
\end{frontmatter}

\section{\label{sec:level1} Introduction }

Handbury-Brown and Twiss introduced intensity interferometry techniques with photons detected by telescopes located at short distance as compared to the size of stars \cite{HBT1956a,HBT1956b}. Since then the HBT  method has found numerous applications in various fields, ranging from quantum optics \cite{SCA2006}, cold atom physics \cite{Nature2005,Cay2020} to nuclear \cite{Koonin1977,Pratt2005} and particle physics \cite{Gold1960,Bia2000}. In nuclear physics, the correlation function (CF), usually at small relative momentum or small angle, has been widely explored to infer the space-time dimension of the emitting source formed in heavy ion collisions (HIC) over a wide energy range. Here, unlike in the original application in astronomy, the temporal evolution of the source, as well as the final state interaction (FSI) between the correlated particles, take effect on the CFs \cite{Koonin1977}. For the review one can refer to \cite{Verde2006,Lisa2005}.
At relativistic energies,  CF  technique has  been developed  to extract the spatial extent of the source \cite{FOPI2005,STAR2001,STAR2010} and to image the anisotropic shape of the fire ball\cite{STAR2004,STAR2005}. Besides  the studies on the space-time characteristics of the relativistic HICs, CFs are  also investigated extensively to deduce the interaction properties of the particle pair. For instance, the interaction between anti-protons are confirmed the same as the protons \cite{STAR2015NAT}. ${\rm \Lambda-\Lambda}$ interaction has been studied via the CFs of ${\rm \Lambda}$ pairs\cite{STAR2015PRL}. An earlier review of the studies on CFs at relativistic energies can be found in \cite{Lisa2005}.
At  Fermi energies, CFs of neutrons, light charged particles (LCPs) and intermediate mass fragments (IMFs) have been extensively measured, too. It has been found that the emission time of IMFs is about 100 fm/c \cite{Brown1997,Kim1991,Bowman1993,Cornell1996}, characterizing the time scale of multifragmentation \cite{Bauge1993,HZY1997}. It turns that the CFs are dependent on the emitting source. In ${\rm ^{129}Xe+^{nat}Sn}$ at 50 MeV/u, the emission time of the fast hot protons from an out-of-equilibrium source is much shorter than that  from an equilibrium source \cite{Gourio2000}. The physical content of the pp (and other like-particle) CFs in the presence of fast dynamical and slow secondary decay emitting sources has been discussed, suggesting that the height and width of the CFs can be used to decompose the different sources  \cite{Verde2002}. CF analysis  of massive fragments reveals the short time scale and dissipative feature of the emission of fragments in the dynamic PLF fission in ${\rm ^{124}Sn+^{64}Ni}$ at 35 MeV/u \cite{Pagano2018}. 


With the ability to image the femtoscopic nuclear system created in HIC at sub-zeptosecond time scale, the HBT method may provide a novel opportunity to explore the isospin dynamics. The transport of the isospin degree of freedom is the effect of the nuclear symmetry energy \esym, which attracts a burst of interest in both astrophysics and nuclear physics for the discovery of the neutron star merging event GW170817 \cite{Oertel2017,Abbott17,Abbott18,De18,Xie2019,ALi2018,NBZhang2018,NBZhang2019,YZ2019}. Despite of great experimental and theoretical progress, \esym~  remains one of the most unknown ingredients in nuclear equation of state (EOS) \cite{Colonna2020,XJ2019,WYJ2020}. More accurate constraint of \esym~ requires further investigations of the fine isospin effect originating  from \esym, for instance, the different transport behavior of neutrons and protons, either free or bounded in light clusters, in heavy ion reactions \cite{WRS2014,ZY2017,Jedele2017}. 

Such investigations require isospin chronology, by which the CF method provides a solution determining  quantitatively the emission time scale of light particles with different $N/Z$. The resolution of the CF method reaches a few fm/c. For instance, at SIS energies, the emission order of p, d, t, $^3$He and $^4$He  has been established by comparing correlations of particles with relative momenta parallel and anti-parallel to the center-of-mass velocity of the unlike particle pairs. It is found that the deduced space-time differences of LCP emitting sources allow two complementary scenarios which are attributed  to the duality of the space and time  undistinguishable from each other \cite{Kotte1999}.  At Fermi energies, by measuring the CFs  of p-p, n-p and p-d in  $^{36}$Ar+$^{27}$Al ($N/Z=1.03$) at 61 \amev, it is shown that for the dynamic emissions, the emission  time  constant satisfies $\tau_{\rm n}<\tau_{\rm d}<\tau_{\rm p}$ \cite{Ghetti2003}.  Besides,  CFs of light particles as well as of IMFs exhibit  the dependence on the initial $N/Z$ of the  reaction system, indicating that \esym~ is playing a role \cite{Ghetti2004,XZG2006,HRJ2007}. 

Despite of such potential ability, using the CF method to quantify  the effect of \esym~  is very difficult. Based on transport model simulations, it has been reported that the stiffness of \esym~  does affect  the CFs of n-p, p-p and n-n pairs \cite{CLW2003}. However,  once the momentum dependent interaction is taken into account, the effect of \esym~ is washed out significantly as in the case of ${\rm ^{52}Ca+^{48}Ca}$ at 80 MeV/u according to the same transport model calculations\cite{CLW2004}, particularly for the less energetic particles. Angular dependence in laboratory of the CFs is reportedly stronger than the isospin effect \cite{Henzl2012}. Even more, the radial expansion and the secondary decay may further  wash out the effect of \esym~ too \cite{Verde2007}.

The above difficulties suggest that the \esym~ is not the only  unknown quantity in heavy ion reactions. It is intriguing to invest further systematic efforts in the studies of the CFs  with an upgraded detection system. Our motivation here is to measure the CFs of  the $Z=1$ isotopes  in a heavier system where the \esym~ effect can be enhanced because of the more neutron-rich environment created with larger space-time extension. In addition, it is also intended to clarify whether the positive d-d correlation peak exists since some previous experimental results have negated the theoretic prediction. In this letter, we present the  CFs of the identical and nonidentical particle pairs of the hydrogen isotopes  in 30 MeV/u \arau~  reactions ($N/Z=1.44$).  An isospin-dependent sequence of particle emission has been confirmed. The Koonin-Pratt (K-P) equation, which connects the CF with the source function and the kernel function as shown by Eq. (3) later in section 3, has been applied to  fit the p-p CF to extract the emission time constant $\tau_{\rm p}$. The positive correlations on d-d and t-t CFs are suggestively presented and discussed.

\section{\label{sec:level2} Experimental setup }

The experiment was performed with the Compact  Spectrometer for Heavy IoN Experiment (CSHINE) installed at the final focal plane  of  the Radioactive Ion Beam Line at Lanzhou (RIBLL-1). The argon beam with 30 MeV/u incident energy was delivered by the Heavy Ion Research Facility at Lanzhou (HIRFL) bombarding on a gold target with the thickness of $1~ {\rm mg/cm^2}$. The first phase setup of CSHINE was installed \cite{Yijie2020}.  The fission fragments (FFs) were measured by three parallel plate avalanche counters (PPACs) with the sensitive area of $240\times280~ {\rm mm^2}$, delivering  the position and the timing information of the coincident FFs. The main PPAC was centered at $\theta_{\rm lab}=50^\circ$, while the other two coincident PPACs were centered at $\theta_{\rm lab}=40^\circ$ and $95^\circ$  on the other side of the beam. The perpendicular distances of the PPACs to the target were equally  $427.5 ~{\rm mm}$ and all the centers of the sensitive areas were in the same horizontal level. The LCPs at midrapidity were measured by two telescopes, each consisting of two silicon strip detectors (SSDs) and one  layer of CsI(Tl) array containing $3\times3$ units read out by photo diode (PD). The first layer of the SSD was a single-sided silicon strip detector (SSSSD) of 32 strips with the thickness of $65~{\rm \mu m}$. The second layer, with the  thickness of $1540~ {\rm \mu m}$,  was a double-sided silicon strip detector (DSSSD) of 32 strips on both sides. The sensitive area $S_{\rm a}$, the distance $d$, the polar angle $\theta_{\rm lab}$ and the azimuth $\phi_{\rm lab}$ of the two SSD telescopes are listed in table I.  Three small silicon telescopes consisting of two Si(Au) barrier detectors and one CsI(Tl) unit was placed at large angles with $\theta_{\rm lab}=90^\circ, 120^\circ$ and $128^\circ$, to measure the evaporated   LCPs. The geometry of detectors were determined by mechanical machining, and the overall accuracy is better than 1 mm.

The correlation functions of LCPs are measured by the SSD telescopes, as schematically displayed in Fig. 1 (a).  The strip widthes of both SSDs are 2mm and the interstrip distance is 0.1 mm. With the pixel size of $2\times2 ~{\rm mm^2}$, the angular resolution of $0.7^\circ$ of the LCPs was achieved in the experiment. The SSDs  were calibrated by combining the precise pulser generator and the $^{239}$Pu $\alpha$  source. The thicknesses of the mylar foil ($\approx 4 \mu m$) and the dead layer ($\approx 0.6 \mu m$ \cite{hira2018}) of SSD detector are corrected. After the SSDs with well-defined thickness were calibrated, the energy deposit in CsI(Tl) units could be calculated using the program LISE++ \cite{Tara2016} and  then calibrated through the  $\Delta E_2-E_{\rm CsI}$ band for each isotope.  For each telescope, the total uncertainty of the  energy of LCPs, which was largely contributed by the CsI(Tl) unit, was estimated to 2\% according to the Monte-Carlo simulation studies reproducing the width of the isotope bands on the $\Delta E_2-E_{\rm CsI}$ plot. The mass resolution of $\Delta M=0.1$ for $\alpha$ particle was then obtained  \cite{Yijie2020}. Fig. 1 (b)  presents the $\Delta E_2-E_{\rm CsI}$ scattering plot of one CsI(Tl) unit in SSD telescope 2. It is shown that the isotopes of $Z=1$ and 2 are clearly identified. Here the non-uniformity of the thin SSSSD is not corrected. Fig. 1 (c) presents the energy spectra of p, d and t.  It is shown that the low energy parts for each isotopes are undetected, because the high voltage for the first layer of the SSD telescopes did not work during the experiment and only the particles penetrating through the second layer  were analyzed here.  
 
\begin{table}[h]
\centering\caption{\label{tab:table1}Geometric parameters of SSD telescope 1 and 2.}
\renewcommand{\arraystretch}{1.1}
\renewcommand{\tabcolsep}{0.7pc}
\begin{tabular}{ccccc}
\hline
  \hline
  SSD Tele.  & $S_{\rm a}$ (mm$^2$) &  $d$ (mm) & $\theta_{\rm lab}~(^\circ)$ &   $\phi_{\rm lab}~(^\circ)$    \\
  \hline
   1 & $64\times64$ &   162 &   51    & 82     \\
   2 &  $64\times64$&   222 &   22    & 64     \\
 \hline
  \hline
\end{tabular}
\end{table}

\begin{figure}[!hbt]
   \centering
   \includegraphics[angle=0,width=0.68\textwidth]{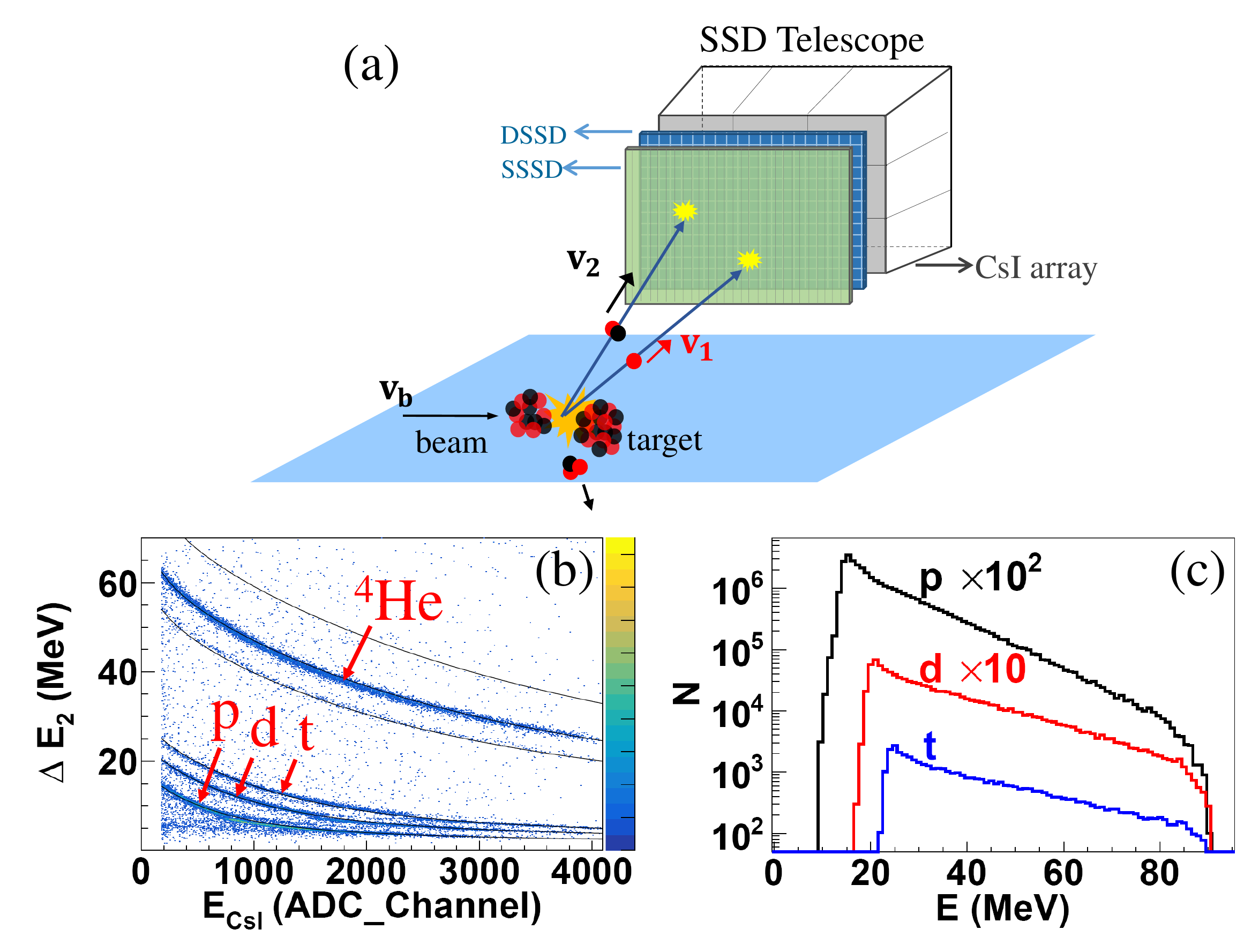}
   \caption{(Color online)   (a) Schematic view of the measurement of two correlated particles from the intermediate velocity source formed in heavy ion reactions using the SSD telescope, (b)   $\Delta E_2-E_{\rm CsI}$ scattering plot of SSD telescope 2, and (c) The energy spectra of p, d and t in the two-body coincidence events.}
\label{setup}
\end{figure}      
    
In the data analysis, we focus on the track recognition and reconstruction in the events with multi hits in SSD telescopes. One first counts the fire multiplicity in the CsI(Tl) array, and then looks for the fired strips crossing the area which matches geometrically to each individual fired CsI(Tl) unit. The pixel of $2 \times 2~ {\rm mm^2}$ at the crossing point of the front strip and the rear strip of the DSSSD defines the hit position. Special check of signal sharing is conducted. In case that two neighboring strips are fired, the energies are summed to form a single hit if the smaller signal is below a given ratio (85\% in the analysis) of the larger one on the neighboring. A good track (hit) on the  DSSSD requires that the relative difference of the signal amplitude between the front  and the rear strip is less than 20\%. Averagely 6\% of the total exclusive events are abandoned because two or more DSSSD hits are recognized in the area corresponding to one single CsI(Tl) unit.  

Considering  the current angular range $20^\circ-60^\circ$ and the high energy threshold in current experiment,  the LCPs are originated mainly from the intermediate velocity source. According to the moving-source fit in previous experiment of the same reaction, where low energy part of the spectra were well covered, the  target-like (TL) evaporation contributes less than $10\%$ in the phase space the SSD telescopes covered \cite{ZY2017,milkau1991}. The projectile-like  (PL) source  may contribute to the CF in the area of large relative momentum if one particle from PL source is measured, see later discussions in Fig. \ref{mixing}. 


\section{\label{sec:level3} Results and Discussions }

Since the event statistics of the 4-body coincidence, i.e.,  2 LCPs in SSD telescopes and 2 FFs in PPACs, does not suffice, only the LCP-LCP two-body coincidence events  are presented.  When two LCPs were identified in the SSD telescopes, the reduced relative momentum ${\textbf{q}}$ can be approximated to the non-relativistic form as 
 
 \begin{center}
\begin{equation}
  {\textbf{q}}=\mu \left( {\textbf{p}}_1/m_1- {\textbf{p}}_2/m_2\right) 
\end{equation}
\end{center}

where $m_i$ and ${\textbf{p}_i}$ represent the mass and momentum of the two particles marked by  $i=1$ and 2, respectively.  $\mu$ is the reduced mass.  The  CF is  constructed by
 
\begin{center}
\begin{equation}
 1+R({q})=C_{\rm N}\frac{Y_{\rm con}({\textbf{p}}_1, {\textbf{p}}_2)}{Y_{\rm mix}({\textbf{p}}_1, {\textbf{p}}_2)}
\end{equation}
\end{center}
where the numerator $Y_{\rm con}$ is the coincident yield  with the particle pair flying with  momenta  ${\textbf{p}}_1$ and ${\textbf{p}}_2$, respectively,  while the denominator $Y_{\rm mix}({\textbf{p}}_1, {\textbf{p}}_2)$ are the yield product of  particle 1 and 2 taken from mixing events in the same data sample. $C_{\rm N}$ is the normalization factor to ensure $R(\textbf{q})=0$ at sufficiently large $q$.

To extract the space-time extension of the source, the correlation function can be fitted using the angle-averaged  Koonin-Pratt equation \cite{Pratt1987} as following

\begin{center}
\begin{equation}
 1+R(q)=1+4\pi\int{S(r,t) \cdot K(r,q) {\rm d} r {\rm d}t} 
\end{equation}
\end{center}

where $S$ is the source function characterizing the space-time extension of the source  and $K$ is  the Kernel function carrying all the information about the wave function of the particle pair and the mutual Coulomb and nuclear FSI, respectively. Applying a Gaussian form to both the size and  the time evolution, the source function is written as

\begin{center}
\begin{equation}
 S(r,t)=c\cdot \exp{\left(-r^2/2\sigma_r^2-t^2/2\tau^2\right)} 
\end{equation}
\end{center}

where $\sigma_r$ is the width of the source size and $\tau$ is the emission time constant.

\begin{figure}[h]
 \centering
 \includegraphics[angle=0,width=0.68\textwidth]{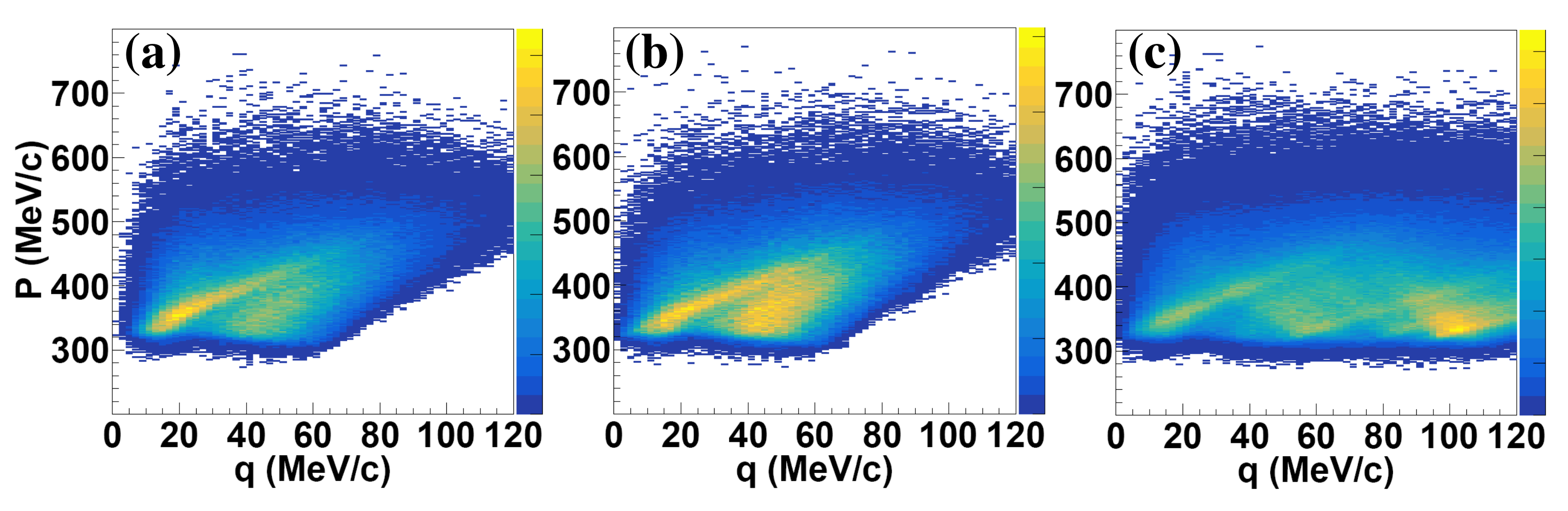}
 \caption{(Color online) (a)  $P-q$ scattering plot of the correlated proton pairs. (b)  $P-q$ scattering plot of the protons picked in mixing events in the same data sample of the two-body events. (c)  $P-q$ scattering plot   of the protons from mixing inclusive events.}
 \label{mixing}
 \end{figure}

The scheme to construct mixing event spectra is essential in the correlation function analysis, as summarized in Fig. \ref{mixing}. Panel (a) presents the scattering plot of the total momentum $P$ and the relative momentum $q$ of the two correlated protons. Two main bands are visible. The upper one, with larger $P$ in clearer coincidence with $q$,  represents the events with two protons recorded in the same SSD telescope,  while the lower one, with smaller $P$ but larger $q$ represents those recorded by the two SSD telescopes, respectively. The light gap between the two bands are due to the dead area between the two telescopes.  Panel (b) presents the $P-q$ scattering plot by randomly mixing the pairs from two successive events in the same  data sample as in panel (a). The distribution exhibits similar feature. Panel (c) presents the  $P-q$ scattering plot by picking the two protons from different events in the inclusive one-body measurement. In comparison with the mixing spectra in panel (b),  the distribution shows more complicated components. An obvious components is seen at very large $q$ but relatively lower total momentum $P$, suggesting that one of the protons may come from a  PL emission which sees higher possibility to be recorded in the inclusive events.  In our analysis, the same data sample is used to  construct the yield product $Y_{\rm mix} ({\textbf{p}}_1, {\textbf{p}}_2)$  in  the denominator of ansatz (2). The validity of the same scheme has been verified in \cite{ghetti2000,elmaani1993}.        
  
\begin{figure}[h]
 \centering
 \includegraphics[angle=0,width=0.68\textwidth]{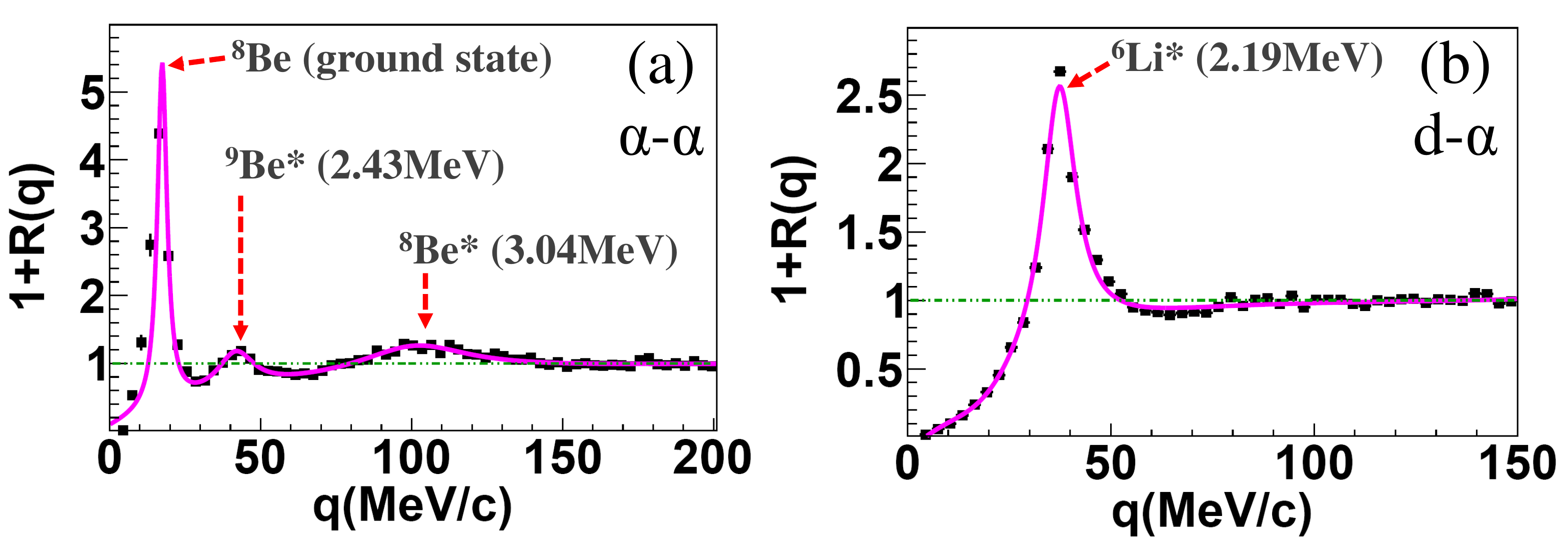}
 \caption{(Color online) $\alpha-\alpha$ (a) and $\d-\alpha$ (b) correlation functions in 30 MeV/u $^{40}$Ar+$^{197}$Au reactions. The individual peaks corresponding to different states are indicated.}
 \label{resonance}
 \end{figure}

 The CFs of  $\alpha$-$\alpha$ and d-$\alpha$ CFs are constructed to verify the scheme of the event mixing and the calibration of the detector, as presented in Fig. \ref{resonance} (a). The three peaks at $q\approx 15,42$ and 100 MeV/c, corresponding to the $^8$Be ground state, the excited states of $E=2.43$ MeV in  $^9$Be  and $E=3.04$ MeV in $^8$Be, are clearly presented. Similarly, the  $q\approx 39$ MeV/c peak, corresponding to 2.19 MeV excited state of  $^6$Li on  d-$\alpha$  CF is presented in Fig. \ref{resonance} (b). The position of the  d-$\alpha$  peak (expected at  $q=42$  MeV/c) is slightly shifted to left by about 1 bin, possibly due to the inaccuracy of the calibration and possibly the collective motion \cite{Verde2007}. The relative yield of the first peak of the $^8$Be ground state is not so enhanced as in literature, because only the $\alpha$ particles at high energy part are measured. Besides, the CF strength $1+R(q)$ situate constantly at unity at large relative momentum $q$, indicating the feasibility of the mixing event scheme, and the energy and momentum conservation induced correlations (EMCIC) may bring insignificant influence \cite{Cha2008}. 


\begin{figure}[h]
 \centering
 \includegraphics[angle=0,width=0.68\textwidth]{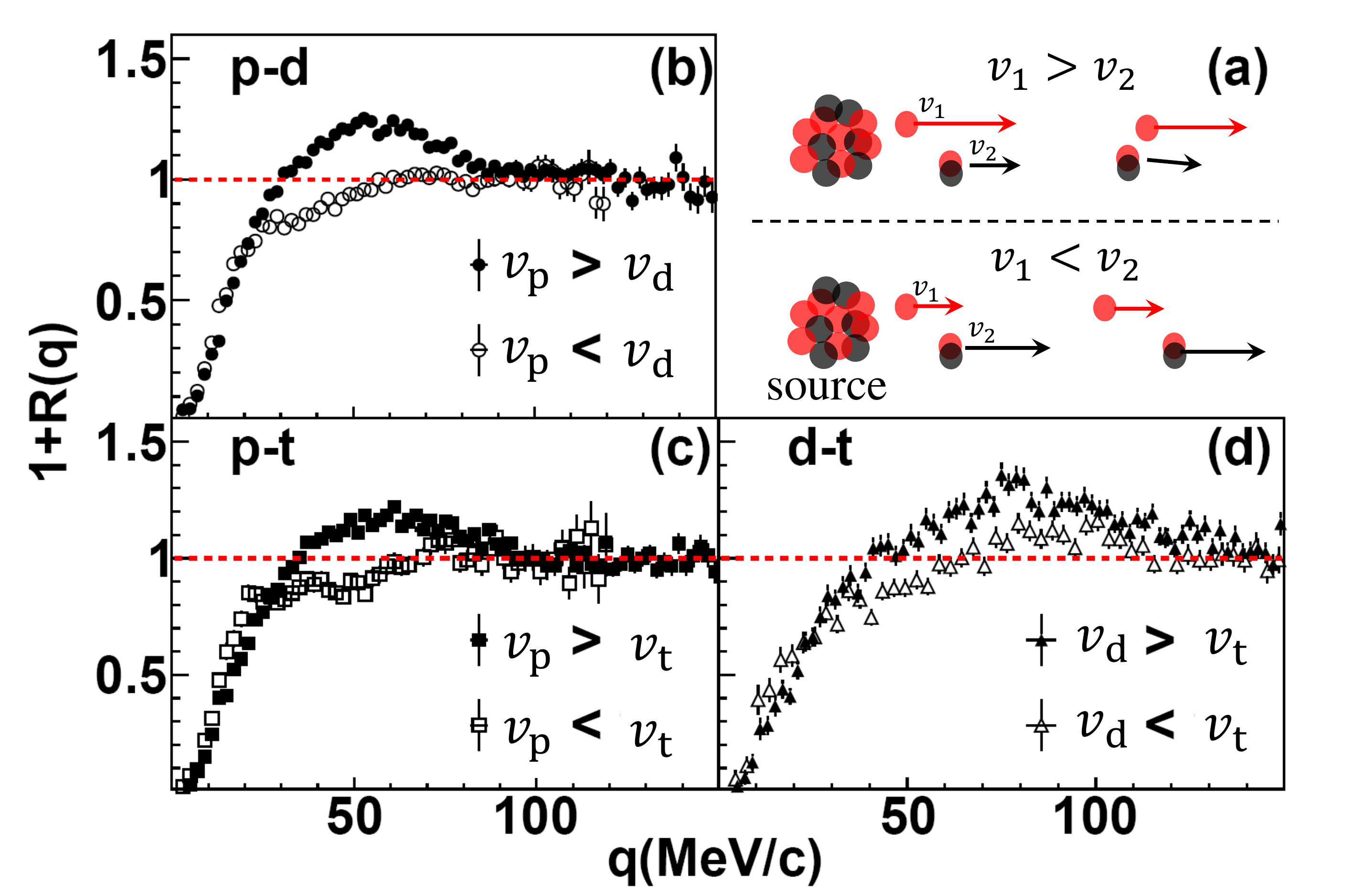}
 \caption{(Color online) (a) The principle of using  the velocity-gated CF to determine the emission order. Here $\tau_1>\tau_2$ is assumed. Panels ($\rm{b-d}$) present the velocity-gated  CFs of p-d, p-t  and  d-t  pairs, respectively. The velocity conditions are indicated in each panel. }
 \label{unlike-pair}
 \end{figure}

We now investigate the emission sequence of the hydrogen isotopes. With less space-time ambiguity, the emission sequence of different particles can be inferred by comparing the CF of the nonidentical particle pair with different velocity gates \cite{Ghetti2001}. Considering two unlike particles marked by 1 and 2 in a classical view, as shown schematically in Fig. \ref{unlike-pair} (a), if the particle 1 is emitted averagely later than particle 2, i.e.,  $\tau_1>\tau_2$,  the (anti)correlation  will be stronger for the group of events with  $v_1>v_2$ where $v$ denotes the velocity,  because particle 1 catches up with particle 2 during the flight and the final-state interaction (FSI) between the particle pair is then expected strong. On the contrary,  the  (anti)correlation  is reduced with  $v_1<v_2$ because the distance between them increases with time. Else if  $\tau_1<\tau_2$  satisfies,  the  (anti)correlation  will be stronger with  $v_1<v_2$.  We note here that the analysis is not done for the 4-body correlations due to the insufficient statistics. It is assumed that different particle pairs do not differentiate the fine division of the reaction centrality.

Fig. \ref{unlike-pair}  ($\rm{b-d}$) present the velocity-gated CFs of p-d,  p-t and d-t  pairs, respectively. The velocity gate conditions are indicated  in each panel.  For p-d pair,  in the group with $v_{\rm p}>v_{\rm d}$ condition, the Coulomb anti-correlation is stronger  at lower $q$  and the positive correlation in the vicinity of $q=50$ MeV/c is shown. In the group of  $v_{\rm p}<v_{\rm d}$, however, the  Coulomb anti-correlation is weaker at lower $q$ and the positive correlation disappears. The discernible difference of the correlation strength between the two groups suggests that  protons are emitted averagely later according to the above  criteria, i.e.,  $\tau_{\rm d}<\tau_{\rm p}$, which is in accordance with earlier experimental result in Ar+Ag at 34 MeV/u \cite{gelderloos95}. Very similar trend can be found  for p-t pair, where the difference between the two CFs is  more pronounced, supporting similarly $\tau_{\rm t}<\tau_{\rm p}$.  For d-t pair, the position of the peak in the vicinity of $q=70$ MeV/c shown in the group of  $v_{\rm d}>v_{\rm t}$ accords with the contribution originating from the loosely bound $^5$He. Similarly, the correlation is stronger in the group of $v_{\rm d}>v_{\rm t}$ and  $\tau_{\rm t}<\tau_{\rm d}$ is correspondingly inferred. Combined, the emission sequence of $\tau_{\rm t}<\tau_{\rm d}<\tau_{\rm p}$  is evidenced in agreement with the order of dynamic emission  $\tau_{\rm n}<\tau_{\rm d}<\tau_{\rm p}$  established in $^{36}$Ar+$^{27}$Al ($N/Z=1.03$) where the neutron emission time constant has also been extracted \cite{Ghetti2003}.

\begin{figure}[h]
 \centering
 \includegraphics[angle=0,width=0.68\textwidth]{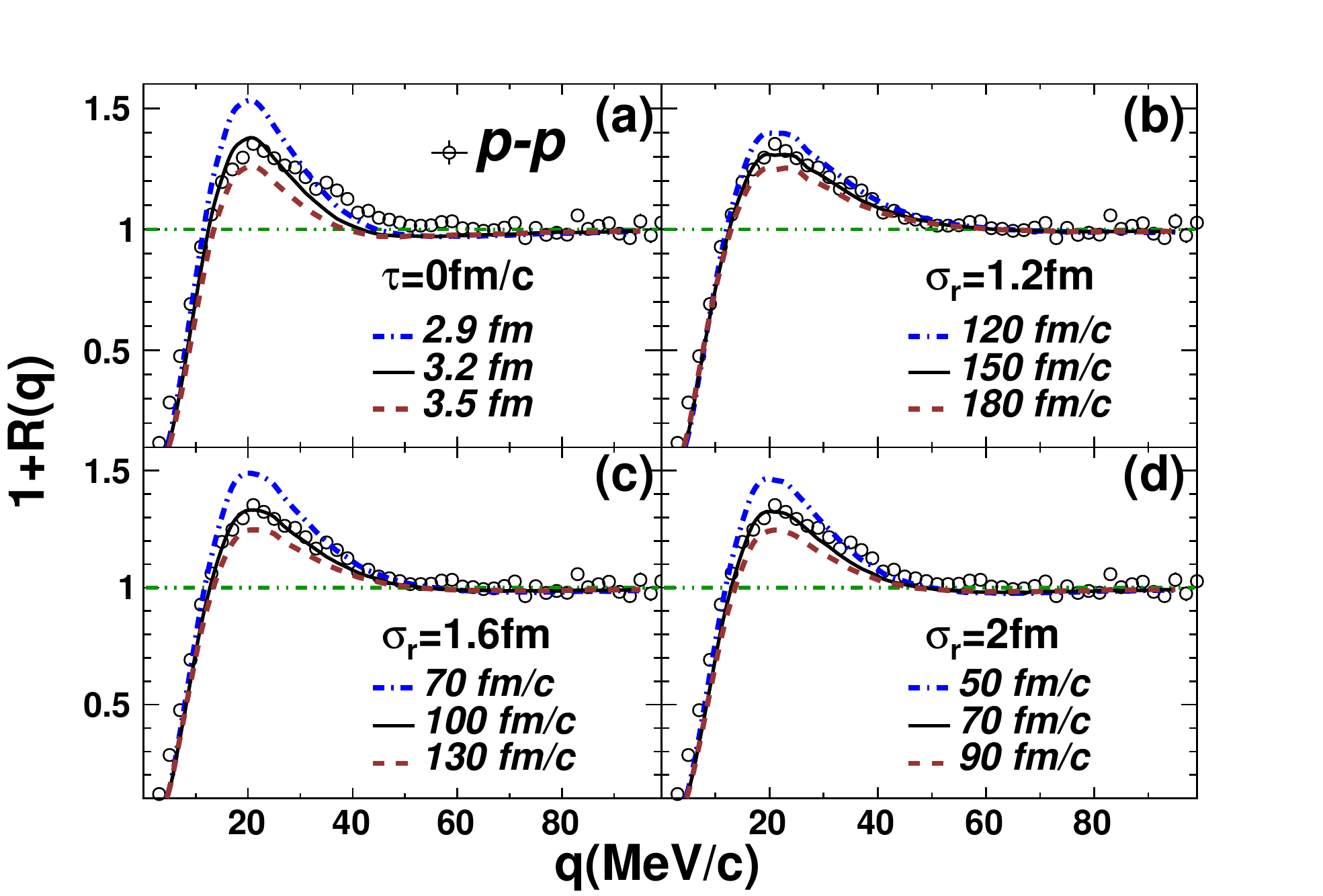}
 \caption{(Color online) The  CF of p-p in comparison with the predictions of CRAB code calculations by varying the emission time constant $\tau_{\rm p}$ with different Gaussian parameter  $\sigma_{\rm r}$ of the spatial extension of the source. Panel (a) presents the calculation results with different source size assuming zero emission time scale. Panels (b-d) present the results by varying $\tau_{\rm p}$  at  $\sigma_{\rm r}=1.2$, 1.6 and 2.0 fm, respectively.}
 \label{pp}
 \end{figure}

The emission time constant of a certain species can be extracted from the  correlation function of the identical particle pair. Fig. \ref{pp} presents the p-p correlation function in comparison with the calculations using Correlation After Burner code (CRAB). The positive correlation peak at $q\approx 20$ MeV/c  is attributed to the S-wave nuclear attraction between the protons. 
It is well  known that  the space-time ambiguity exists since the source size $r$ and the time constant of the particle emission $\tau$ are correlated \cite{Glasm1994}. Such ambiguity can be discriminated partly by the directional cut analysis on the relative velocity \cite{Lisa1993} . But the statistics here does not allow us for the fine analysis. So before inferring the space-time extension of the source, which is assumed to be the intermediate velocity source (midrapidity source) since the high energy protons are mainly measured in the experiment, we first estimate the size of the source containing the transferred projectile nucleons and the target nucleons of the same amount. The mass number of the source reads

\begin{center}
\begin{equation}
 A_{\rm s}=2\times A_{\rm p} \times LMT
\end{equation}
\end{center}
 where LMT is the linear momentum transfer. Assuming  the most probable momentum transfer of 75\%, one writes $A_{\rm s}=60$ and the static radius is about $r=4.8$ fm. 
 
Fig. \ref{pp} (a) first presents the comparison of the p-p CF to CRAB calculation \cite{Pratt1994,Crabweb} by fixing the emission time constant at $\tau=0$. The fit with a Gaussian standard width $\sigma_{\rm r}=3.2$ fm reproduces the height of the CF but the shape is slightly off.  Then, if releasing the condition of $\tau=0$, one expects to obtain a smaller source size and a finite emission time. Panels (b-d) presents the results by varying the $\tau$ at given geometric size parameter $\sigma_{\rm r}=1.2$, 1.6 and 2.0 fm, respectively. As shown, at all $\sigma_{\rm r}$ settings, the calculation can reproduce the experimental CF reasonably well. Surveying more carefully, it is found that at $\sigma_{\rm r}=2.0$ fm, the shape of fit starts to deviate from the   right descent of the experimental CF although the height coincide with the data points. With a Gaussian standard width $\sigma_{\rm r}=1.6$ fm  the best  fit situates at $\tau_{\rm p}=100$ fm/c.  An uncertainty of $\pm50$ fm/c is allowed if one takes into account the variation of the size parameter. The value of $\tau_{\rm p}$ are consistent with the prediction of transport model simulations based on coalescence  scheme of clustering \cite{LWNPA}. In addition, the value of $\tau_{\rm p}$ is comparable to the time that the projectile traverses the target region, suggesting again that the contributions from PL  and TL evaporation are insignificant because these two sources experience mainly evaporation process characterized by much larger time scale. We recall here that  ansatz (4) is a simplified way to characterize the space-time extension of the source. The possible mixing of different emitting source and the collective motion of the participant zone may brings some influence to the modeling of the source, but the magnitude of $\tau$ is unlikely changed dramatically. Alternatively, the imaging technique extracts the source profile by a numerical inversion of K-P equation without introducing \emph {a priori} assumptions on the source shape \cite{Verde2002}. 

\begin{figure}[h]
 \centering
 \includegraphics[angle=0,width=0.68\textwidth]{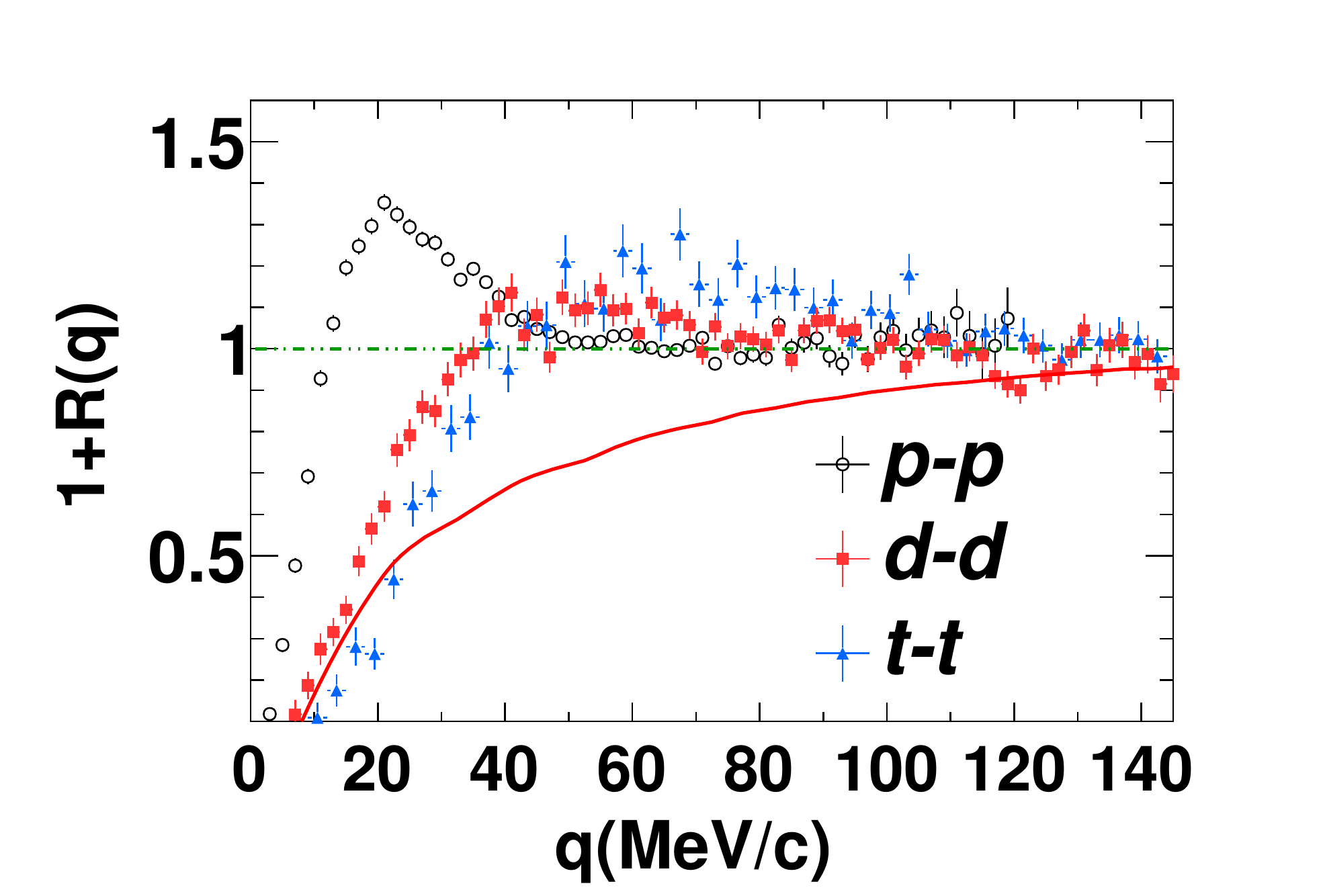}
 \caption{(Color online)   The  CFs of p-p, d-d and t-t, respectively. The red thin curve is the CRAB calculation of d-d correlation using the source-size  parameter $\sigma_{\rm r}$ =1.6fm, while the energy (momentum) distribution of deuterons are derived from the slope of   its energy spectrum.}
 \label{pp-dd-tt}
 \end{figure}

Fig. \ref{pp-dd-tt}  presents the experimental  CFs of d-d and t-t in comparison to p-p. It is clear that the Coulomb anti-correlation at small $q$, visible for all pairs, increases successively from proton to deuteron and triton.  Given that the Coulomb force between the particle pair is the same for p-p, d-d and t-t, which are all $Z=1$ isotopes, the evolution of the strength of the Coulomb anti-correlation further is consistent with the particle emission hierarchy that the neutron-rich triton are emitted with an averagely shorter time constant than that of deuteron and proton, in agreement with the observation of Fig. \ref{pp}. The increasing anti-correlation from p-p to d-d and t-t pairs is consistent with the experimental observation in $^{40}$Ar+Ag at lower beam energy, where faster emission of t and d was also suggested \cite{elmaani94}. One shall keep cautious that the above inference from Fig. \ref{pp-dd-tt} relies on the assumption that p, d and t emissions are characterized by the same geometrical source size. A more convincing picture is foreseeable if matrices of analysis can be done with the relative momentum $\textbf{q}$ being perpendicular and parallel to the sum momentum $\textbf{P}$ of the source \cite{Lisa1993}.

 Another  feature worth noticing in Fig. \ref{pp-dd-tt} is the positive correlation peaks near 45 (60) MeV/c on the d-d (t-t) correlation function slightly beyond the statistical uncertainty. This is at variance with the experimental results in earlier literature \cite{Chitwood1985,Gelder1995,Poch87}, where the positive correlation was not observed possibly because of the low granularity of the detector. In addition, the CRAB calculation does not reproduce the positive correlation if a simplified Woods-Saxon d-d potential is adopted, as shown by the curve in Fig.\ref{pp-dd-tt} for which the source-size parameter is used as $\sigma_{\rm r} =1.6$ fm and the temperature is obtained from the deuteron energy spectrum.  On the other hand, however, the presence of the positive correlation is qualitatively consistent with the theoretic prediction of the resonating group  (RG) method  by taking into account the form factor of deuteron nucleus \cite{Chitwood1985,Chwieroth1972}. Interestingly, It is noticed that the positive correlations have also been experimentally indicated, although not discussed, at the same positions of the d-d and t-t CFs in  $^{40}$Ar+Ag at 17 MeV/u \cite{elmaani94}.
 Comparing to the t-t correlation, which exhibits an even more pronounced positive peak, the d-d positive correlation is weaker. If the presence of the positive peak in d-d and t-t CFs is true, it suggests that the nuclear FSI is possibly playing a role, and that the weaker d-d positive correlation is consistent with the effect of the formation of $\alpha$ from two deuterons at small distance in phase space. We checked the possible effect of the cluster recognition algorithms  in the neighboring strips, as well as the $\alpha-\alpha$  correlation function, which did not show such wide positive correlation except for the known peaks corresponding to the $^8$Be and $^9$Be states \cite{Yijie2020}, the positive correlation could not be excluded. Further studies are called for to understand the presence of the positive peaks in d-d and t-t CFs.

\begin{figure}[h]
 \centering
 \includegraphics[angle=0,width=0.68\textwidth]{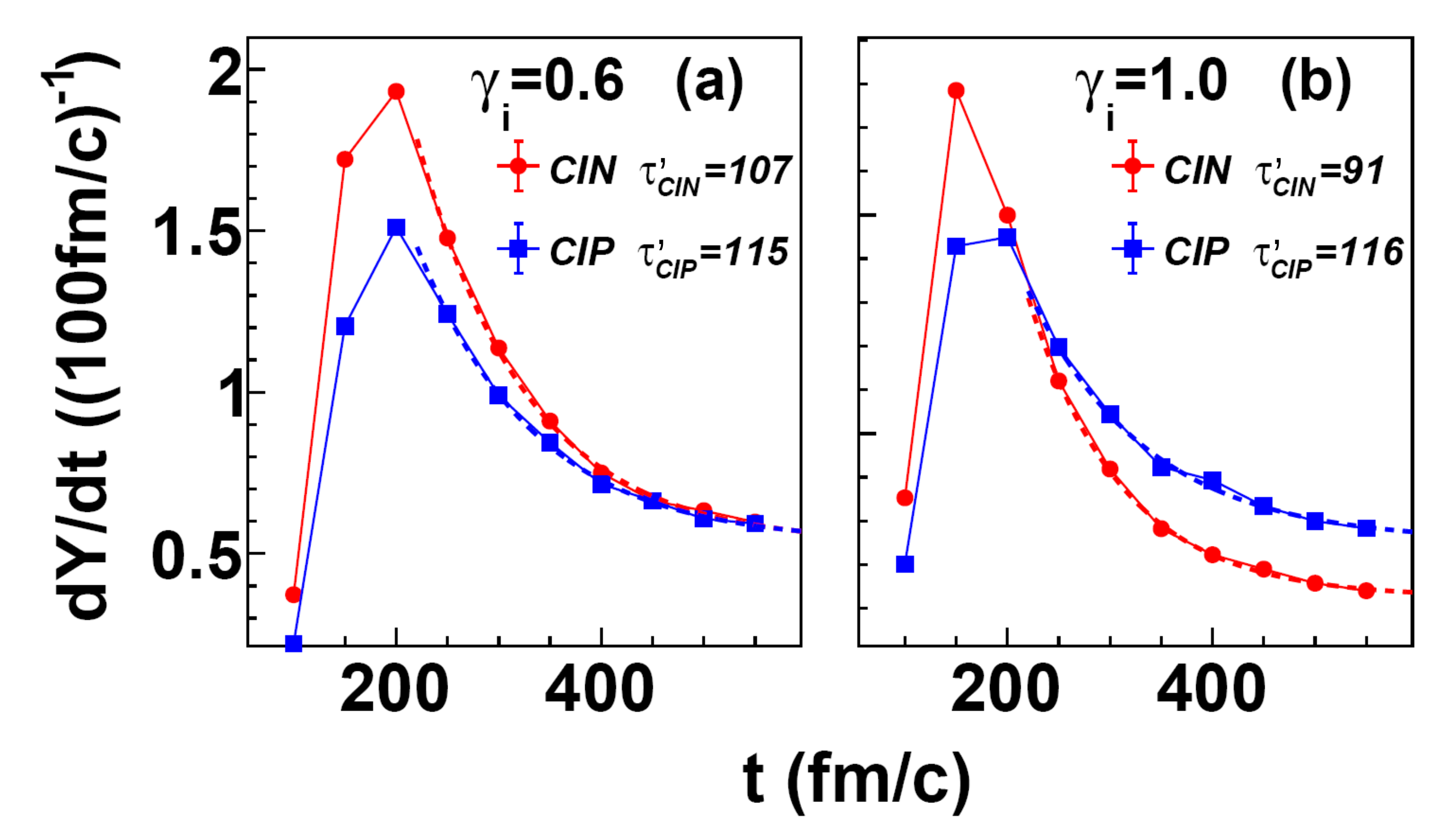}
 \caption{(Color online)  The emission rate of CIN (red) and CIP (blue) as a function of time in ImQMD  simulation of 30 MeV/u Ar+Au by averaging the impact parameter over $b=1-5$ fm. Two slope parameters of \esym, $\gamma_{\rm i}=0.6$ (a) and 1.0  (b), are adopted representing a soft and a stiff \esym, respectively. The dashed curves are the exponential fit to the decaying points with the time constants $\tau^\prime_{\rm CIN}$ and $\tau^\prime_{\rm CIP}$ all listed.}
 \label{fig7}
 \end{figure}

The observed emission sequence of $Z=1$ isotopes, i.e.,  a N/Z effect on the CFs, may or may not be an evidence that the \esym~ is at work. Different emitting  source, secondary decay, collective motions \cite{Verde2002,Henzl2012,Verde2007}, as well as the different transport behavior of neutrons and protons, all possibly cause the isospin-dependent hierarchy. More efforts are required to extract convincing isospin effect from the isotope-resolved CFs. As a starting attempt, we  inspect the relaxation of the isospin degree of freedom (IDOF) with transport model calculations since we are measuring the energetic LCPs  in the current phase space insignificantly contributed by the evaporation according to previous moving source analysis\cite{ZY2017,milkau1991}. 
In an intuitive picture, neutrons (protons) experience  repulsion (attraction) due to the isovector nuclear potential, thus neutrons are driven more easily to the gas phase causing isospin fractionation \cite{XHS2000}. With a stiffer \esym, the neutron-proton difference in terms of emission time constant is larger.

It is of interest to reproduce the IDOF relaxation process  with an Improved Quantum Molecular Dynamics (ImQMD) model  incorporating the density dependence of nuclear symmetry energy \esym. Instead of pursuing a quantitative comparison with the experimental emission time constant, we aim at revealing qualitatively the characteristic of the IDOF relaxation by inspecting the time distribution of the emission of coalescence invariant neutrons (CINs) and  protons (CIPs). Here CINs (CIPs) refers to the total neutron (proton) numbers in the light particles with $Z<3$. The details of the ImQMD calculations with IQ3 interaction potential for the same reactions can be found in \cite{WQH2015,WQH2020}.  Fig. 7 presents the time evolution of the emission rate of CINs and CIPs during the dynamic emission stage within $t<600$ fm/c. Here $t=100$ fm/c corresponds to the time when the projectile and the target start to contact on surface. To see the effect of \esym, two types of density dependence with the slope parameter $\gamma_{\rm i}=0.6$ (a) and 1.0 (b) are simulated, representing a soft and a stiff \esym~ respectively. The energy cut of $E_{\rm k}/A>10$ MeV is applied approximately as in the experiment. The calculation is done by weighting  the impact parameter in the range  $b=1-5$ fm. It is shown that after the production peak, the emission rate of CINs descends more rapidly than CIPs, suggesting neutrons experience a faster dynamic emission than protons, either free or bounded in clusters. We fit the descent with an exponential decay function ${\rm exp}(-t/\tau^\prime)$, where the superscript $^\prime$ is written to differentiate it from the Gaussian extension in Eq. (4). For $\gamma_{\rm i}=1.0$, the time difference $\tau^\prime_{\rm CIP}-\tau^\prime_{\rm CIN}$ is larger than that at $\gamma_{\rm i}=0.6$, indicating a faster IDOF relaxation with stiffer \esym~ in agreement with the picture obtained from the angular distribution of the $N/Z$ of  LCPs in the same reaction\cite{ZY2017}. The qualitative agreement of ImQMD simulation with experiment suggests that the precise measurement of isospin chronology in the whole phase space, including neutrons which have not been measured here, raises further scientific interest of experimental and theoretical endeavors on the way towards comprehending the effect of \esym~ on CFs in heavy ion reactions.

\section{Summary}
In summary, with the Phase-I CSHINE detector appended to RIBLL-1, we have measured  the  CFs of like and unlike particle pairs of hydrogen isotopes in \arau~ at 30 \amev.  The velocity-gated CFs of p-d, p-t and d-t pairs support the emission sequence of $\tau_{\rm p}>\tau_{\rm d}>\tau_{\rm t}$.  For the identical particle pairs, the CF of p-p is reproduced by K-P equation with $\tau_{\rm p}\approx100$ fm/c. Besides, the anti-correlation at small $q$ becomes increasingly  pronounced from p-p to d-d and t-t pairs, consistent with the emission hierarchy that the neutron-rich particles are emitted relatively faster under the assumption that p, d and t emissions are characterized by the same spatial parameter. A positive peak is observed on both d-d and t-t  CFs at variance with earlier experiments but in qualitative agreement with the theoretic prediction of resonating group method, requiring further confirmation in future experiments. The quantitative determination of the isospin chronology can be potentially used to probe the isospin dynamics. Nevertheless, such task is not yet straightforward since various conditions  intertwine in the reaction process, including beam energy, reaction geometry, collective motion and the convolution of different emitting sources etc. In addition, neutrons carrying the first-order isospin effect need to be detected with high accuracy. Therefore further efforts, both theoretically and experimentally, are called for to better understand the effect of  \esym~ at $\rho\le\rho_0$ on intensity interferometry in heavy ion reactions. 

\section{Acknowledgment}  
This work has been supported by the National Natural Science Foundation of China under Grant Nos. 11875174, 11890712, 11961131010 and U1732135 and by Ministry of Science and Technology under Grant No. 2020YFE0202001. The work is also supported   by the Initiative Scientific Research Program and the Center of High Performance Computing of Tsinghua University and by Heavy Ion Research Facility at Lanzhou (HIRFL). The authors acknowledge the machine staff for delivering the beam.

\end{document}